\documentclass[reprint, superscriptaddress, amsmath,amssymb, aps, prl]{revtex4-2}
\usepackage{graphicx}

\usepackage{silence}
\usepackage{epstopdf}
\usepackage{physics}
\usepackage{color,times}
\usepackage{bm}
\usepackage{dcolumn}
\usepackage{multirow}
\usepackage{amssymb,amsfonts,amsmath,graphicx}
\usepackage{epsfig}
\usepackage{xcolor}
\usepackage{SIunits}
\usepackage{braket}
\usepackage{epstopdf}
\usepackage{physics}
\usepackage{stmaryrd}
\usepackage{bbold}
\usepackage{amsmath}
\usepackage[utf8]{inputenc}
\usepackage{chngcntr}
\usepackage[english]{babel}
\usepackage[colorlinks=true, breaklinks=true, linkcolor=blue, citecolor=blue, urlcolor=blue]{hyperref}
\usepackage{textcomp}
\usepackage{soul}
\usepackage{makecell}

\newcommand{\mb}{\mathbf}

\definecolor{teal_custom}{rgb}{0, 0.6, 0.53}

\newcommand{\ParisAddress}{Universit\'e Paris-Saclay, Institut d'Optique Graduate School,\\
	CNRS, Laboratoire Charles Fabry, 91127 Palaiseau Cedex, France}
\newcommand{\HarvardPhysicsAddress}{Department of Physics, Harvard University, Cambridge, Massachusetts 02138, USA}
\newcommand{\OviedoAddress}{Nanomaterials and Nanotechnology Research Center (CINN-CSIC), 
	Universidad de Oviedo (UO), Principado de Asturias, 33940 El Entrego, Spain}
\newcommand{\CASAddress}{Institute of Physics, Chinese Academy of Sciences, Beijing 100190, China}

\newcommand{\BerkeleyPhysicsAddress}{Department of Physics, University of California, Berkeley, CA 94720, USA}
\newcommand{\LBNLAddress}{Material Science Division, Lawrence Berkeley National Laboratory, Berkeley, CA 94720, USA}
\newcommand{\CMUAddress}{Department of Physics, Carnegie Mellon University, Pittsburgh, PA 15213, USA}

\newcommand{\MBIAddress}{Max-Born-Institut, Max-Born-Strasse 2A, 12489 Berlin, Germany}

\newcommand{\PrincetonAddress}{Department of Electrical and Computer Engineering, Princeton University, Princeton, NJ 08544, USA}

\begin{document}

\author{Guillaume Bornet$^*$}
\affiliation{\ParisAddress}
\affiliation{\PrincetonAddress}

\author{Marcus Bintz$^*$}
\affiliation{\HarvardPhysicsAddress}

\author{Cheng Chen$^*$}
\affiliation{\ParisAddress}
\affiliation{\CASAddress}

\author{Gabriel Emperauger$^*$}
\affiliation{\ParisAddress}
\affiliation{\MBIAddress}

\author{Mu Qiao}
\affiliation{\ParisAddress}

\author{Romain Martin}
\affiliation{\ParisAddress}

\author{Daniel Barredo}
\affiliation{\ParisAddress}
\affiliation{\OviedoAddress}

\author{Shubhayu~Chatterjee}
\affiliation{\CMUAddress}

\author{Vincent~S.~Liu}
\affiliation{\HarvardPhysicsAddress}

\author{Thierry~Lahaye}
\affiliation{\ParisAddress}

\author{Michael~P.~Zaletel}
\affiliation{\BerkeleyPhysicsAddress}
\affiliation{\LBNLAddress}

\author{Norman~Y.~Yao}
\affiliation{\HarvardPhysicsAddress}

\author{Antoine~Browaeys}
\affiliation{\ParisAddress}

\title{Dirac Spin Liquid Candidate in a Rydberg Quantum Simulator}

\begin{abstract}
    We experimentally investigate a frustrated spin-exchange antiferromagnet in a quantum simulator, composed of $N=114$ dipolar Rydberg atoms arranged into a kagome array.
    Motivated by a recent theoretical proposal of a gapless $U(1)$ Dirac spin liquid ground state, we use local addressing to adiabatically prepare low-energy states.
    We measure the local polarization and spin-spin correlations over this adiabatic protocol, and observe our system move from a staggered product state, through an intermediate magnetic crystal, and finally into a disordered, correlated liquid.
    We estimate the entropy density of this atomic liquid to be similar to that of frustrated magnetic insulators at liquid nitrogen temperatures.
    We compare the correlations in our liquid to those of a simple, parameter-free \textit{ansatz} for the Dirac spin liquid, and find good agreement in the sign structure and spatial decay.
    Finally, we probe the static susceptibility of our system to a local field perturbation and to a geometrical distortion.
    Our results establish Rydberg atom arrays as a promising platform for the preparation and microscopic characterization of quantum spin liquid candidates.
\end{abstract}

\maketitle

The spin-$1/2$ kagome antiferromagnet is the paradigmatic two-dimensional setting in which geometric frustration suppresses classical order and amplifies quantum fluctuations~\cite{elserNuclearAntiferromagnetismRegistered1989}.
In its canonical form---nearest-neighbor antiferromagnetic Heisenberg exchange on corner-sharing triangles---a broad body of both experimental and theoretical work indicates a pronounced absence of conventional symmetry breaking~\cite{misguichTWODIMENSIONALQUANTUMANTIFERROMAGNETS2013,chenKagomeHeisenbergAntiferromagnet2025}.
The resulting quantum-disordered state seems to persist across deformations of the canonical model, including further-neighbor exchange and easy-plane anisotropies~\cite{huVariationalMonteCarlo2015, zhuChiralCriticalSpin2015, heDistinctSpinLiquids2015, lauchliQuantumSimulationsMade2015}.
However, the precise nature of this phase is still actively debated.
Competing theoretical pictures include fractionalized gapped topological spin liquids~\cite{sachdevKagomeTriangularlatticeHeisenberg1992, yangPossibleSpinliquidStates1993, Han2012}, as well as gapless liquids with Dirac-like low-energy structure~\cite{hastingsDiracStructureRVB2000, ranProjectedWaveFunctionStudySpin2007, heSignaturesDiracCones2017}.
The gapless case is particularly intriguing in that a Dirac spin liquid
constitutes an effective realization of fermionic quantum electrodynamics in 2+1 dimensions; this provides a route from a microscopic lattice Hamiltonian to a strongly-interacting relativistic field theory 
via emergence~\cite{Hermele2008}.

These theoretical ambiguities are paralleled in experiments on kagome-lattice and other frustrated magnetic insulators.
By their nature, spin liquids do not come with a simple bulk order parameter; correspondingly, the absence of magnetic order is not, by itself, a positive experimental identification.
Instead, the distinguishing experimental signatures are expected to appear in the structure of correlations and excitations---including dynamical responses and, more subtly, entanglement-related diagnostics and multi-spin observables~\cite{kitaevTopologicalEntanglementEntropy2006, gregorDiagnosingDeconfinementTopological2011, bondersonMeasuringTopologicalOrder2021}.
The gapless spin liquid scenario is, in principle, more favorable in that it should exhibit universal algebraically decaying spin correlations and power-law, low-temperature thermodynamics.
In realistic settings, however, even these signatures are readily obscured by chemical disorder, structural imperfections, and additional dynamical degrees of freedom such as phonons~\cite{huangHeatTransportHerbertsmithite2021, chamorroChemistryQuantumSpin2021, murayamaUniversalScalingSpecific2022}.
Thus, candidate kagome spin liquid materials are often in limbo: evidently lacking conventional order, yet not sharply distinguished from a trivial paramagnet~\cite{wenExperimentalIdentificationQuantum2019}.

\begin{figure}
	\centering
	\includegraphics{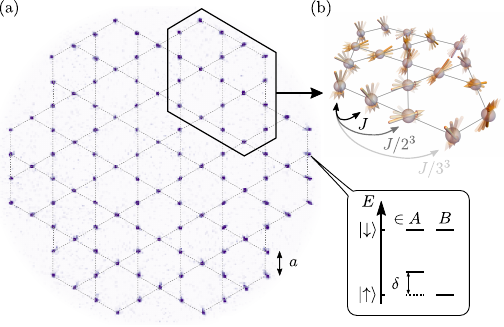}
	\caption{\textbf{Dipolar XY model on a kagome array.}
		(a)~Fluorescence image of $N = 114~$individual $^{87}\text{Rb}~$atoms 
		trapped in a kagome array of optical tweezers (dotted lines emphasizing the kagome structure are guides to the eye). The lattice spacing is $a = 12\,\mu$m.
		(b)~Schematic depicting the dipolar XY model. The spin states are encoded in a pair of Rydberg states which exhibit dipolar spin-exchange interactions. We denote $A$ the  sub-array of the addressed atoms whose $\ket{\uparrow}$ state is lightshifted by $\delta$ and $B$ the sub-array of the non-addressed atoms.
	}
	\label{fig:fig1}
\end{figure}

%
The last few decades have seen the advent of a complementary quantum-simulation-based approach, where frustrated magnets are assembled from individually controlled quantum particles~\cite{duanControllingSpinExchange2003,santosAtomicQuantumGases2004, altmanQuantumSimulatorsArchitectures2021}.
Indeed, magnetic elements, polar molecules, Rydberg atoms and atomic defects all possess local moments that can interact through strong antiferromagnetic dipolar coupling~\cite{hazzardQuantumCorrelationsEntanglement2014},  spiritually similar to the antiferromagnetic exchange of solid state materials. 
More recently, these experiments have entered a new frontier in which low entropy densities can be achieved in frustrated systems of hundreds of spins~\cite{Scholl2021,mongkolkiattichaiQuantumGasMicroscopy2023, xuFrustrationDopinginducedMagnetism2023,  guoSiteresolvedTwodimensionalQuantum2024}; this enables novel, microscopic probes of highly-entangled magnetic states that are difficult to obtain in solids.
Related capabilities have enabled  the targeted preparation of several known topological wavefunctions and observations consistent with beyond-Landau phenomena ~\cite{everedProbingKitaevHoneycomb2025, iqbalNonAbelianTopologicalOrder2024, iqbalQutritToricCode2025,Semeghini2021, deleseleucObservationSymmetryprotectedTopological2019}.

%
In this Letter, we study the many-body physics emerging from frustrated dipolar spin-exchange interactions in a Rydberg atom simulator. 
Motivated by theoretical predictions of a gapless Dirac spin liquid ground state~\cite{Bintz2024}, our experimental setup consists of a two-dimensional, $N=114$ site kagome array of $^{87}\text{Rb}$ atoms trapped in optical tweezers (Fig.~\ref{fig:fig1}a). 
We rely on strong resonant dipole-dipole interactions~\footnote{ $\mu = |\langle{60S} | e \vec{r}_e |{60P}\rangle | \approx 4.5$ kilodebye---about 2400 times greater than the electric dipole moment of a water molecule.} between two electronic Rydberg states of opposite parities.
We encode the pseudo-spin 1/2 on the states $\ket{\uparrow} = \ket{60S_{1/2}, m_J = 1/2}$ and $\ket{\downarrow} = \ket{60P_{1/2}, m_J = -1/2}$, leading to an effective easy-plane spin-exchange Hamiltonian,
\begin{equation}
    H_{\mathrm{XY}} = \frac{\hbar J}{2} \sum_{i<j} \frac{a^3}{r_{ij}^3}(\sigma^x_i \sigma^x_j + \sigma^y_i \sigma^y_j)
\end{equation}
where $\sigma ^{x,y}$ are spin-1/2 Pauli matrices, $r_{ij}$ is the distance between atoms $i$ and $j$, the lattice spacing is $a=12$ $\mu$m, and at this spacing the measured exchange coupling strength is $J/(2\pi)=0.77$ MHz.
We apply a 45 G magnetic field perpendicular to the plane of the array to ensure spatially isotropic interaction~\cite{ravetsMeasurementAngularDependence2015, emperaugerTomonagaLuttingerLiquidBehavior2025}.

Our main results are three-fold. First, we access the low-energy many-body physics of $H_{\rm XY}$ by dynamically controlling a spatially staggered effective magnetic field along $z$: an initial product state at large field quasi-adiabatically follows toward a strongly entangled phase as the field is ramped down. 
The prepared state exhibits diffuse 
spin–spin correlations at high lattice momenta, consistent with the absence of magnetic order; moreover, four-body bond–bond correlations suggest that it is not a valence-bond solid~\citep{Anderson1987, Baskaran1987}. 
By reversing the adiabatic ramp, we estimate the  entropy density of this kagome spin liquid candidate to be 0.6 $\ln$ 2 per site.
Second, we compare the measured correlations to those of a simple \textit{ansatz} wavefunction for a Dirac spin liquid, and find remarkable similarities. Finally, we measure the local static susceptibility of the system to a single-site pinning field, and to geometric deformation.
%


\begin{figure*}
	\centering
	\includegraphics
    {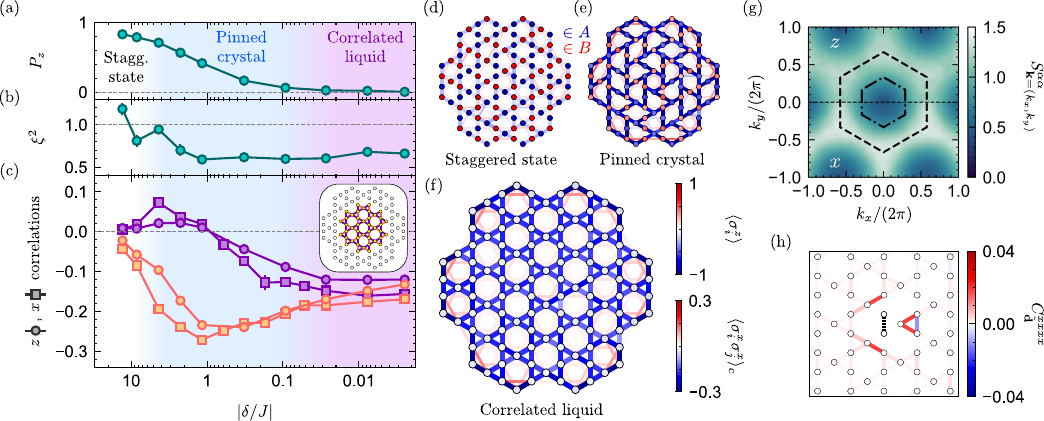}	\caption{\textbf{Adiabatic preparation of the ground state.} 
        (a)~Average staggered $z$-magnetization $P_z$ as a function of $|\delta / J|$.
        (b)~Evolution of the entanglement witness $\xi^2$ (see text) as a function of $|\delta/J|$.
		(c)~Average nearest-neighbor connected correlations measured along $z$ (circle markers) 
		and $x$ (square markers) as a function of $|\delta / J|$. 
		To avoid edge effects, the purple and light blue data show the average calculated using only the purple and light blue nearest-neighbor pairs shown in the inset. We identify three phases: the staggered state, 
		the pinned crystal phase and the  correlated liquid phase. 
		The background colors are a guide to the eye to represent these different phases.
		(d-f)~For each phase ($|\delta / J| = [13.0,1.16,0.0025]$), we represent the $z$ magnetization $\langle \sigma^z_i \rangle$ 
		of each atom (colored circles) and the connected nearest-neighbor and next-nearest-neighbor correlations 
		$\langle \sigma^z_i\sigma^z_j \rangle_c$ (colored bonds). 
		In the staggered initial state, the atoms with a positive magnetization belong to the sub-array $A$, and the others belong to $B$. 
        (g)~Momentum-space structure factor of the correlated liquid in the $\sigma^z$ (top) and $\sigma^x$ (bottom) bases. The dashed and dash-dot line represent the extended Brillouin and regular Brillouin zone~\cite{supp}.
        (h)~Four-body bond-bond correlations (see text). 
	}
	\label{fig:fig2}
\end{figure*}

\textit{Adiabatic preparation of a correlated liquid.}---To prepare the ground state of $H_{\rm XY}$ we follow an adiabatic procedure similar to that used in our previous works~\cite{Chen2023, emperaugerTomonagaLuttingerLiquidBehavior2025}. 
The experiment starts by initializing a classical antiferromagnet along $z$, that is, a staggered arrangement of spins $\ket{\downarrow}\in A$ and $\ket{\uparrow}\in B$ with  total magnetization $\langle M^z \rangle = \langle \sum_{i}\sigma^z_i \rangle = 0$.
The design of this staggered state is chosen to have maximal rotational symmetry, uniform distribution of the conserved $z$ magnetization, and local similarity to a pattern suggested by theoretical analysis~\cite{Bintz2024, supp}.
We create this state in the presence of focused addressing laser beams, which produce a staggered effective magnetic field, $H_{\rm Z} = \hbar\delta\sum_{i\in A}(1+\sigma^z_i)/2$ where $A$ is the sub-array of the addressed atoms [Fig.~\ref{fig:fig2}(d)], the $B$ sub-array is not addressed, and $\delta$ is the amplitude of the applied lightshift.
Our initial state is then approximately in the ground state of the total Hamiltonian $H_{\rm{XY}}+H_{\rm{Z}}$ for $\delta \gg J$, with a per-site preparation fidelity of about 93\%~\cite{supp} (not including detection error).

We then adiabatically ramp down the light shift as $\delta(t) = \delta_0 e^{-t/\tau}$ (taking $\tau = 0.3~\mu$s and $\delta_0/(2\pi) = 10~$MHz) to connect the staggered state to low-energy states of $H_{XY}+ H_Z$, for varying $\delta$.
Finally, we readout the state of the atoms by destructive, total-system snapshot measurements in either the $\sigma^x$ or $\sigma^z$ basis~\cite{supp}.
This sequence is repeated up to $\sim 30,000~$times in order to compute the magnetizations and spin correlations with good statistical accuracy.

The results of this procedure are shown in Fig.~\ref{fig:fig2}.
As we reduce the applied field, the spins gradually lose their staggered magnetization, $P_z=(\langle \sigma^z\rangle_B - \langle \sigma^z\rangle_A)/2$, [Fig.~\ref{fig:fig2}(a)].
Simultaneously, they develop correlations $\langle \sigma^\alpha_i \sigma^\alpha_j\rangle_c = \langle \sigma^\alpha_i\sigma^\alpha_j\rangle - \langle\sigma^\alpha_i \rangle\langle\sigma^\alpha_j\rangle$ in both $\alpha=(x,z)$ bases
\footnote{We have checked experimentally that, owing to the $U(1)$-symmetry, the $\sigma^y$ measurements are identical to the $\sigma^x$ ones \cite{supp}.}.
In fact, the summed correlations are already indicative of nontrivial entanglement forming in the system, as computed by e.g. the entanglement witness
$\xi^2 =  \frac{1}{2N} \left[ {\rm Var}(M^{x}) +{\rm Var}(M^{y})+ {\rm Var}(M^{z}) \right]< 1$ ~\cite{Toth2009} [Fig.~\ref{fig:fig2}(b)].
Our local readout allows us to picture these quantum correlations in much greater detail [Fig.~\ref{fig:fig2}(c-f)].
At very large field, the system remains close to a product state  [Fig.~\ref{fig:fig2}(d)].
Then, as $\delta$ decreases, neighboring atoms on opposite subarrays begin to exchange their spins and develop strong antiferromagnetic correlations, while atoms within the same subarray remain weakly correlated.
This gives the state a crystal-like appearance, with a geometry pinned by the applied field [Fig.~\ref{fig:fig2}(e)].
Averaging these correlations over nearest-neighbor pairs (restricting to the bulk 42 sites, to avoid edge effects) we observe that this pinned crystal pattern is most distinct for $\delta/J\approx 1$ [Fig.~\ref{fig:fig2}(c)].
As $\delta$ is further reduced, $\delta/J<0.01$, the local correlations smooth out into a much more homogeneous configuration [Fig.~\ref{fig:fig2}(c,f)]. 
In the remainder, we focus on the detailed characterization of this final, liquid-like state.

First, we probe the longer-range correlations in momentum space, by examining the equal-time spin structure factor, $S^{\alpha\alpha}(\mathbf{k}) = \sum_{i,j} e^{i \mathbf{k }\cdot \mathbf{r}_{ij}}\langle \sigma^\alpha_i \sigma^\alpha_j\rangle_c$.
This is the quantity typically probed in 
neutron scattering investigations of antiferromagnetic solids.
As shown in Fig.~\ref{fig:fig2}(g), both the $\sigma^x$ and $\sigma^z$ structure factors exhibit diffuse weight around the perimeter of the extended Brillouin zone~\cite{supp}.
We conclude there is a lack of magnetic order in the system.
Moreover, we note a similarity between the $\sigma^x$ and $\sigma^z$ structure factors, suggestive of the 
emergence of an enhanced $SU(2)$ spin symmetry.

Next, we examine whether nearest-neighbor pairs have a tendency to form spatially structured correlations, i.e. valence bond solids, in which each atom pairs with one of its neighbors into a singlet, breaking lattice translation and rotational symmetry.
To do so, we compute the four-body (Ursell connected~\cite{kuboGeneralizedCumulantExpansion1962}) correlation function, $\expval{\sigma^\alpha_i\sigma^\alpha_j\sigma^\alpha_k\sigma^\alpha_l}_c$.
This probes the likelihood of the nearest-neighbor spin pair $(k,l)$ to become anticorrelated, if the nearest-neighbor pair $(i,j)$ is.
We average over combinations of nearest-neighbor bonds separated by the same displacement $\tilde{\mathbf{d}}$ and get $C^{\alpha\alpha\alpha\alpha}_{\tilde{\mathbf{d}}}$, taking $(i,j)$ to be in the center 42 sites [orange sites shown in the inset of Fig.~\ref{fig:fig2}(c)], while $(k,l)$ is allowed to range over the full array (the details of these calculations are given in~\cite{supp}).
The results are depicted in Fig.`\ref{fig:fig2}(h): we observe statistical attraction and repulsion between valence bonds at short distances, but which rapidly decay.
This lack of long-range solid order indicates our state is not a simple statistical mixture of a few valence bond solids.

\begin{figure}
	\centering
	\includegraphics{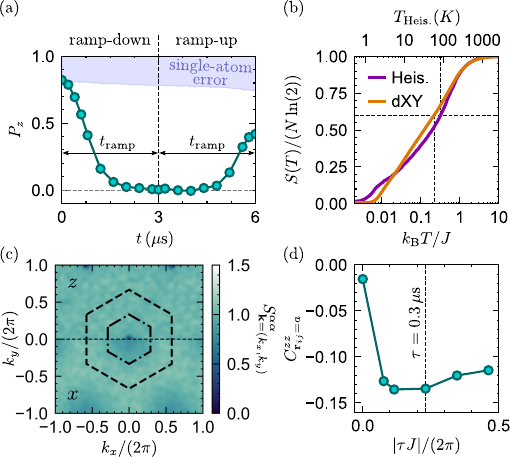}
	\caption{\textbf{Thermometry}. (a)~Staggered magnetization dynamics during a mirrored forward-then-backward ramp. 
    (b) Numerical calibration curve of thermodynamic entropy as a function of temperature. Orange line: dipolar XY model on an $N=24$ site torus. Purple line: $N=42$ nearest-neighbor Heisenberg model (adapted from Ref.~\cite{schnackMagnetism42Kagome2018}). Upper axis $T_{\text{Heis.}}$ is the equivalent absolute temperature for Herbertsmithite. 
    (c)~Spin structure factor following a rapid quench to a high-temperature state. (d)~Nearest-neighbor $\sigma^z$ correlations for varying ramp times $\tau$.
    }
    \label{fig:fig3}
\end{figure}

To summarize, Fig.\ref{fig:fig2}(f,g) show that the prepared state has strong short-range correlations and lacks symmetry-breaking magnetic order, and the four-body correlations
 indicate no obvious affinity to  valence-bond solid order. We will hence designate it as a \textit{correlated liquid}.
\textit{Thermometry.}---How hot is our liquid?
To assess this, we use the system as its own thermometer, in the $\delta\gg J$ paramagnetic regime: there, single-body physics dominates and the partition function approximately factorizes. 
Thermodynamic quantities are then simple functions of the proportion, $p=(1+P_z)/2$, of spins aligned with the applied staggered field, e.g. the entropy density, $S/N = -p \ln p - (1 - p) \ln(1 - p)$.
For the correlated liquid, we assume its thermodynamic entropy is the sum of (1) that of the initial product state, $S(0)=0.3 \ln 2$, and  (2) the entropy produced by effective heating during the ramp (i.e. from diabaticity and decoherence).
To determine the heating, we perform an experiment in which we ramp $\delta$ for $t = 3\mu$s, as before, and then immediately run the ramp in reverse [Fig.~\ref{fig:fig3}(a)]. 
We measure the returned staggered magnetization to be $P_z = 0.42$, corresponding to $S(2 t_\text{ramp}) = 0.9 \ln(2)$ (correcting for detection errors).
Assuming the forward and backward ramps produce the same entropy, we estimate the entropy of the correlated liquid as the average of these two points, $S(t_\text{ramp})/N = 0.6 \ln 2$.
Finally, this entropy can be calibrated against numerical thermometry curves~\cite{schnackMagnetism42Kagome2018}, shown in Fig.~\ref{fig:fig3}(b).
These indicate that (i) our correlated liquid is at $k_BT \approx 0.2 \hbar J$, and (ii) for the paradigmatic nearest-neighbor kagome Heisenberg material, Herbertsmithite~\cite{Han2012}, this entropy density corresponds to temperatures at the liquid nitrogen scale (80 K)~\footnote{The entropy density of our initial state ($S/N=0.9\ln 2$) corresponds to the liquid helium temperatures used for solid-state inelastic neutron scattering measurements.}.


%
There are several sources of entropy in our experiment.
First, single-body errors in the staggered state preparation lead to the initial non-unity polarization, which accounts for half of the estimated total entropy in the correlated liquid.
Second, the ramp takes place in finite time, so unitary diabatic errors are potentially important.
Indeed, we checked that rapidly quenching (in less than $100\,$ns) the staggered field from its initial value to zero produced a high-temperature state, with significantly weaker correlations [Fig.~\ref{fig:fig3}(c)]. 
Third, Rydberg atoms have a finite lifetime, $T\sim100\,\mu$s~\cite{supp}.
Combined with other imperfections, such as positional and motional disorder of the atoms, this lifetime limits the length of the ramp. 
We probed this tradeoff by varying the decay time $\tau$ of our ramp: measuring at time $t=10\tau$, so $\delta/\delta_0\approx e^{-10}$, we observed that the nearest-neighbor correlations have a maximum amplitude when using ramps with $\tau\approx 0.3\,\mu$s [Fig.~\ref{fig:fig3}(d)].
Future experiments, we expect, will be able to access lower temperatures by using higher Rydberg states, cryogenic chambers, and increased laser power~\cite{nguyenQuantumSimulationCircular2018, schymikSingleAtoms6000Second2021, zhangHighOpticalAccess2025, jinExtendedRydbergLifetimes2026}.
%

\begin{figure}
	\includegraphics{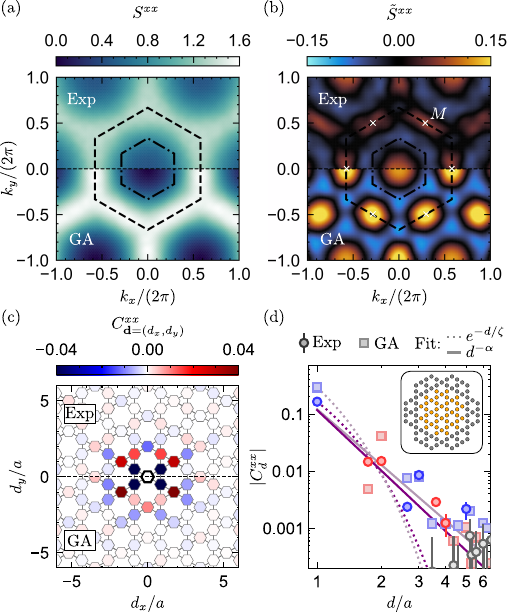}
    \caption{\textbf{Liquid correlations.}
    (a)~Spin structure factor in the experimentally prepared liquid (top) and theoretical Gutzwiller \textit{ansatz} (bottom).
    (b)~Modified structure factor, with the nearest-neighbor correlations subtracted off.
    (c)~Map of correlation function $C^{xx}_{\mathbf{d} = (d_x,d_y)}$.
    (d)~Spatial decay of correlations, comparing the experiment (solid circles) to the Gutzwiller \textit{ansatz} (light squares). We average over pairs $(i,j)$ where $i$ is in the center 42 sites (inset) and $j$ is anywhere in the array. Blue markers indicate $C^x_d<0$, red markers indicate $C^x_d>0$, and gray markers are data within statistical error of zero. The results of fits are the following: $\alpha_{\rm GA} = -3.5 \pm 0.7$, $\alpha_{\rm Exp} = -4.0 \pm 1.0$, $\zeta_{\rm GA}/a = 0.33 \pm 0.03$ and $\zeta_{\rm Exp}/a= 0.34 \pm 0.02$.
    }
\label{fig:ansatz}
\end{figure}

\textit{Comparison to algebraic liquid.}---Despite the present imperfections, could our correlated liquid still exhibit some nontrivial features?
As a guide, we choose to reference against a simple, parameter-free \textit{ansatz} wavefunction for a kagome lattice quantum spin liquid.
In particular, we take it to be the gapless $U(1)$ Dirac spin liquid, which serves as a quantum-critical parent state for many competing orders on the kagome lattice (symmetry-breaking and topological~\cite{hermelePropertiesAlgebraicSpin2008, songUnifyingDescriptionCompeting2019}); a numerical work involving some of us argued this liquid was likely the ground state of $H_{\mathrm{XY}}$ \cite{Bintz2024}.
The \textit{ansatz} is constructed by the standard method of Gutzwiller-projecting a free-fermionic Slater determinant wavefunction into the single-occupancy spin Hilbert space~\cite{ranProjectedWaveFunctionStudySpin2007, beccaQuantumMonteCarlo2017, supp}.
In Fig.~\ref{fig:ansatz}, we compare the spin-spin correlations of the experiment to the theoretical liquid \textit{ansatz}. 
The structure factor of either is dominated by diffuse weight at the edge of the extended Brillouin zone [Fig.~\ref{fig:ansatz}(a)].
In fact, most of the weight comes from nearest-neighbor correlations, washing out finer features.
Making use of our real-space readout, we can subtract this component off, computing  $\tilde{S}^{xx}(\mathbf{k}) = \sum'_{i,j} e^{i\mathbf{k}\cdot \mathbf{r}_{ij}} \langle \sigma^x_i \sigma^x_j \rangle_c $.
Here, the restricted sum $\sum'$ is taken over pairs $(i,j)$ with $r_{ij}>a$, and $i$ in the 42-atom bulk of the array.
As shown in Fig.~\ref{fig:ansatz}(b), these longer-distance correlations are remarkably similar between the experiment and the \textit{ansatz}. 
Notably, both have enhanced weight at the $M$ points of the extended Brillouin zone.
These wavevectors correspond to a 120$^{\circ}$ coplanar order, which analytically and numerically is expected to have large quantum fluctuations in the Dirac spin liquid~\cite{Zhu2019, Zhang2020, Bintz2024}.
%

%
We then examine the spin correlations in real space.
The field theory of the Dirac spin liquid predicts that such correlations decay asymptotically as a sum of oscillating power-laws, $C_{ij} \sim \sum_\alpha C_a \cos(\mb{r}_{ij}\cdot \mb{k}_\alpha) / r_{ij}^{2\Delta_\alpha}$, with universal critical exponents and oscillation wavevectors, $\Delta_\alpha$ and $\mb{k}_\alpha$, and non-universal coefficients, $C_\alpha$, where $\alpha$ labels different emergent excitations.
For example, the $\mb{k}=M$ correlations seen in Fig.~\ref{fig:ansatz}(b) can theoretically be generated by monopoles, which have $\Delta\approx1.3$~\cite{songUnifyingDescriptionCompeting2019, karthikScalingDimension42024}.
Cleanly resolving these component power-laws requires measuring very weak, long-distance correlations that are beyond the present capabilities of atomic platforms.
For that reason, we instead simply compare the (possibly non-universal) short-to-intermediate distance behavior between the experiment and the \textit{ansatz}~\footnote{Whether the \textit{ansatz} has asymptotic correlations that match the field theory is also unknown.}.

Fig.~\ref{fig:ansatz}(c) displays the two-body correlation function averaged over pairs $(i,j)$ with the same displacement $\mb{d}=\mb{r}_j-\mb{r}_i$ (up to a global lattice rotation~\cite{supp}), with $i$ in the bulk and $j$ anywhere in the array.  
The ferromagnetic/antiferromagnetic sign structure of the correlations agrees well between the experiment and the \textit{ansatz} wavefunction. 
To compare the magnitudes, we average over pairs at the same distance $d$, and plot the absolute value $|C^{xx}_d|$ in Fig.~\ref{fig:ansatz}(d). 
The envelopes of both the experiment and the \textit{ansatz} exhibit a fast spatial decay; quantitatively, a power-law fit $|C^{xx}_d|\sim d^{-\alpha}$ gives $\alpha_{\rm{Exp}}=3.5\pm0.7$ and $\alpha_{\rm{GA}} = 4.0\pm 1.0$.   
This is substantially more rapid than what we previously observed in the unfrustrated ground states of $H_{\rm{XY}}$ on the square lattice and the linear chain~\cite{Chen2023, emperaugerTomonagaLuttingerLiquidBehavior2025}.

%

%
In the Supplemental Material, we additionally examine how the correlations are modified near the array's edge~\cite{supp}.
We also compute the correlations of a gapped $Z_2$ spin liquid candidate state~\cite{lu2SpinLiquids2011, yangFrustratedResonatingValence2012, luUnificationBosonicFermionic2017}.
While it is difficult to distinguish between the exponential decay of the $Z_2$ ansatz and the power-law decay of the $U(1)$ DSL on the $N=114$ cluster, we find that the sign of the experimental next-nearest-neighbor ($d=\sqrt{3}$) correlation is in agreement with the $U(1)$ ansatz, and opposite to the $Z_2$. 
However, both disagree with the experiment regarding the sign of the (small) cross-hexagon $d = 2$ correlations.
We have also used linear spin-wave theory to compute the finite-size, $T=0$ correlation functions of the $q=0$ anti-ferromagnetic order, which predicts a magnetization plateau of $|C^{xx}| = 0.3$, one hundred times higher than the $d\ge5$ experimental correlations.
Comparison with finite-$T$ calculations for various symmetry breaking orders would be interesting for future work.

\textit{Static susceptibilities.}---We conclude by measuring the sensitivity of our correlated liquid to Hamiltonian perturbations.
In the solid-state, such static susceptibility measurements are indisposable for material characterization. 
Hence, we here make use of the local controllability of our quantum simulator to adiabatically prepare (from the same initial staggered configuration) the ground state of a perturbed Hamiltonian, $H_{\rm{XY}}+H'$, and examine how it differs from the unperturbed liquid.

\begin{figure}
	\includegraphics{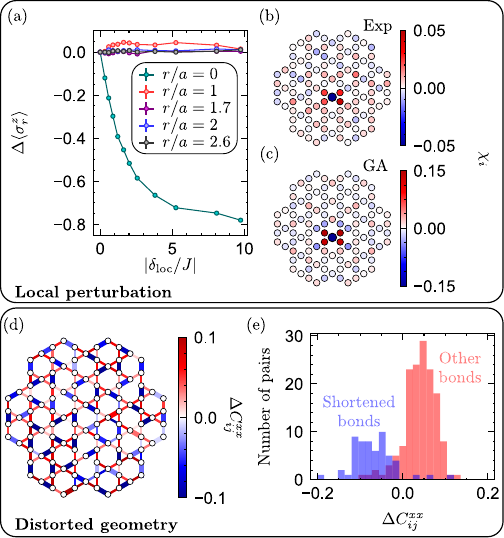}
    \caption{\textbf{Static susceptibilites.}
    (a)~Change in magnetization $\Delta \langle \sigma^z_r\rangle = \langle \sigma^z_r\rangle(\delta_{\rm{loc}}) - \langle \sigma^z_r\rangle(\delta_{\rm{loc}} = 0)$ induced by a single-site field, $\delta_{\rm{loc}} \sigma^z_0$ as a function $\delta$, averaged over sites at distance $r$ from the targeted atom.
    (b)~Local map of the experimentally measured susceptibility. The position of the atom addressed with the local addressing beam is indicated by a slightly larger circle.
    (c)~Theoretical expectation from Gutzwiller \textit{ansatz}.
    (d)~Change in correlations $\Delta C^{xx}$ due to lattice distortion.  Bonds shortened by 5\% are thickened.
    (e)~Histogram of distortion response on nearest-neighbor bonds.
}
\label{fig:static}
\end{figure}

For our first example, we apply a static field to a single atom, $H'=\delta_{\rm{loc}} \sigma^z_0$, by using an additional laser to lightshift its $\ket{\uparrow}$ state.
This follows theoretical proposals to probe emergent spinons by detecting their Friedel oscillations off of such an impurity potential~\cite{dallatorreFriedelOscillationsProbe2016, Bintz2024}.
We prepare liquids in the presence of varying field strengths, $\delta_{\rm{loc}}$, and measure the resulting $\sigma^z$ magnetization of the targeted atom, as well as all other sites in the array.
As shown in Fig.~\ref{fig:static}(a), the targeted spin gradually becomes aligned with the local field.
Due to total $M_z$ conservation, the remaining sites must on average polarize in the opposite direction; we find that this predominantly occurs on the nearest-neighboring sites, with a very weak response at distances greater than one lattice spacing.
Focusing on the low-field regime, we extract the local susceptibility $\chi_{i} = -d \langle \sigma^z_i\rangle /d(\delta_{\rm{loc}}/J)$ at $|\delta_{\rm{loc}}/J| = 0$. 
The resulting response function is shown in Fig.~\ref{fig:static}(b), and the expectation from the \textit{ansatz} is shown for comparison in Fig.~\ref{fig:static}(c). 
Neither shows clear evidence of the expected spinon Friedel oscillations, indicating that our present measurement is limited not only by experimental imperfections, but also by finite-size effects~\cite{supp}.
Still, we hope that similar tests will be theoretically proposed and experimentally attempted in the near future.
%

%
Second, we distort the underlying geometry of the atom array so as to energetically favor a valence bond solid~\cite{seifertSpinPeierlsInstabilityU12024}.
In particular, we select 57, non-overlapping nearest-neighbor pairs, each composed of one atom in sublattice $A$ and the other in sublattice $B$.
As we observed already, such pairs easily develop strong antiferromagnetic correlations at intermediate $\delta$.
By shortening the distance in these pairs by  5$\%$ (0.6~$\mu$m), which enhances the interaction strength along those bonds by about $17$$\%$, we emulate a worst-case example of positional disorder.
We thus expect strong correlations on these shortened bonds and reduced correlations elsewhere.
Accordingly, we measure $C^{xx}_{ij}$ in the presence of this distortion and compare it to the undistorted case, $C^{xx}_{ij,0}$, plotting $\Delta C^{xx}_{ij} = C^{xx}_{ij}-C^{xx}_{ij,0}$ in Fig.~\ref{fig:static}(d).
With the exception of a few outliers, the shortened bonds indeed become more antiferromagnetic while the remaining nearest-neighbor bonds typically do the opposite; the mean and standard deviation of these shifts are $\Delta C^{xx}_{\rm{short}} = -0.069 \pm 0.049$ and $\Delta C^{xx}_{\rm{other}}= +0.041 \pm 0.036$ [Fig.~\ref{fig:static}(e)].

Interestingly, longer-distance correlations are not substantially affected by this perturbation---whether this robustness persists to lower temperatures and larger system sizes remains an open question~\cite{supp}.
%

In summary, we have prepared and characterized low-energy quantum states of a frustrated, exchange-interacting  antiferromagnet of cold atoms. 
Our system does not appear to be ordered, and its correlations are comparable to theoretical expectations for a quantum spin liquid.
The Rydberg array platform is rapidly developing along several directions, which will soon open broad new avenues for additional study of this atomic spin liquid candidate. 
\textit{Note added.}---After submission of our paper, we became aware of two related work which explore non-equilibrium $U(1)$ spin liquids in Rydberg quantum simulators~\cite{geim2026, Karch_inprep_U1QSL}.


\textit{Data availability.} The data is publicly available on \href{https://zenodo.org/records/18638413?token=eyJhbGciOiJIUzUxMiJ9.eyJpZCI6ImIwNGIyNDlmLWMyMTctNDcwNi1hZWMzLTkxYzRkMDdhN2ZiMyIsImRhdGEiOnt9LCJyYW5kb20iOiJhMjkwNTE3OGRhYWEyY2RkOTVmMzk3ZGJhNzVkYWI5OCJ9.m2sTZqlpm_K3T29Zk9REqmNSWBoiujQ11E5Y0w3GgnxMIYoOjoG2HlUFhN_SjuPmDYGAr0yMcJMBWYdnddjI-Q}{zenodo}~\cite{bintz2026}. We encourage those interested to find their own spin liquid observables inside it~\footnote{We offer a nice French dinner to anyone who finds something interesting!}.

\begin{acknowledgments}
    We acknowledge fruitful discussions with Bastien G\'ely,  Yuki Torii Chew and  Lukas Klein.
	This work is supported by the Agence Nationale de la Recherche (ANR-22-PETQ-0004 France 2030, project QuBitAF), and the European Research Council (Advanced grant No. 101018511-ATARAXIA), and the Horizon Europe programme HORIZON-CL4- 2022-QUANTUM-02-SGA (project 101113690 PASQuanS2.1),
    DB acknowledges support from MCIN/AEI/10.13039/501100011033 (PID2024-162669NB-I00).
	MZ was supported by the U.S. Department of Energy, Office of Science, Basic Energy Sciences, under Early Career Award No. DE-SC0022716.
\end{acknowledgments}

\bibliography{references_kagome, footnotes_and_other}
\newpage

\setcounter{figure}{0}
\renewcommand\thefigure{S\arabic{figure}} 

\appendix

\section{Experimental methods}\label{SM:Exp_details}

The implementation of the dipolar XY Hamiltonian relies on the $^{87}\text{Rb}$ Rydberg-atom tweezer array platform described in previous works~\cite{Chen2023,Bornet2024}.
The tweezer array is created by diffracting a 820-nm laser beam on a first Spatial Light Modulator (SLM) and focusing it by a $\text{NA}=0.5$ aspherical lens~\cite{Nogrette2014}. 

We encode our spin states as $\ket{\uparrow} = \ket{60S_{1/2}, m_J = 1/2}$ and $\ket{\downarrow} = \ket{60P_{1/2}, m_J = -1/2}$. We use microwave at $\omega_0/(2\pi)\approx16.4~$GHz to manipulate the spin states. We apply a $45$-G quantization magnetic field perpendicular to the array to ensure isotropic XY interactions and shift away the irrelevant Zeeman sublevels from the $\ket{\uparrow} \leftrightarrow \ket{\downarrow}$ transition. 
The microwaves used to manipulate the qubit states are generated using a vector signal generator (R\&S\textsuperscript{\textregistered}SMM100A) plugged to an antenna placed outside the vacuum chamber.

\subsection{Experimental sequence}\label{SubSM:sequence}

\begin{figure}
	\includegraphics{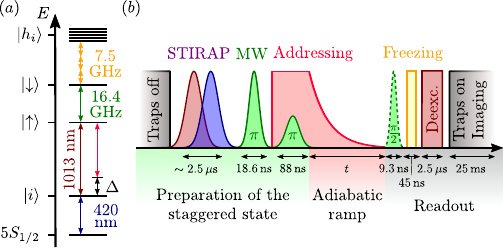}
\caption{\textbf{Experimental sequence.} (a) Atomic energy levels involved in this work and associated transitions: 420 nm and 1013 nm are the wavelengths used for Rydberg excitation, and 16.4 GHz and 7.5 GHz are microwave fields for driving Rydberg-Rydberg transitions. (b) Sketch of the experimental sequence for the adiabatic preparation of the ground state. For clarity the timings of the pulses are not to scale.
}
\label{fig:exp_seq}
\end{figure}

\subsubsection{Assembled array of Rydberg atoms}\label{SubSubSM:Rydberg_arrays}

Figure~\ref{fig:exp_seq} shows the detailed experimental sequence used to prepare the ground state of the kagome array. After randomly loading atoms into optical tweezers of $\approx1~$mK depth (with a typical filling fraction of $60\%$), we assemble the array. We then cool the atoms to a temperature of $\approx 5\,\mu$K using Raman sideband cooling and optically pump them to $\ket{g}=\ket{5S_{1/2}, F = 2, m_F = 2}$. Following this, we adiabatically ramp down the power of the trapping light to reduce the tweezer depth by a factor $\sim 4$. Then, we switch off the tweeezers and excite the atoms to the Rydberg state $\ket{\uparrow}$. The excitation is performed by a two-photon stimulated Raman adiabatic passage (STIRAP) with $420$-nm and $1013$-nm lasers.

\subsubsection{Staggered state preparation}\label{SubSubSM:Staggered_state_prep}

Prior to the adiabatic procedure, we prepare the staggered state $\ket{\uparrow\downarrow\uparrow\downarrow\cdots}$. 
After having initialized all the atoms in $\ket{\uparrow}$, we transfer them from $\ket{\uparrow}$ to $\ket{\downarrow}$ using a microwave $\pi$-pulse 
with a gaussian temporal profile $\Omega(t) = \Omega_{0}e^{-\pi \left( t/t_{\pi} \right)^2}$ taking $\Omega_{\max}/(2\pi) = 26.9~$MHz and $t_{\pi} = 18.6~$ns. 
We then turn on the addressing beam, applying a lightshift of $\delta_0/(2\pi) = 10~$MHz on the atoms in sub-array $A$. We then transfer back the atoms in $B$ from $\ket{\downarrow}$ to $\ket{\uparrow}$ using an other microwave $\pi$-pulse (with $\Omega_{0}/(2\pi) = 7.0~$MHz and $t_{\pi} = 88.0~$ns) while the atoms belonging to $A$ remain in $\ket{\downarrow}$.

\subsubsection{Readout}\label{SubSubSM:Readout}

To avoid residual interactions between the atoms during the readout procedure, we first apply a microwave pulse at $7.5~\text{GHz}$ to transfer the atomic population from $\ket{\downarrow}$ to the $n=58$ hydrogenic manifold $(h_i)$ via a three-photon transition. The atoms in $(h_i)$ do not interact with those remaining in $\ket{\uparrow}$, effectively freezing the interaction dynamics. The next step consists of a deexcitation pulse achieved by applying a $2.5~\mu\text{s}$ laser pulse on resonance with the transition between $\ket{\uparrow}$ and the short-lived intermediate state $6P_{3/2}$ (denoted by $\ket{i}$ in the following), from which the atoms decay back to the ground state $5S_{1/2}$. Finally, we ramp back on the trapping lights, recapturing only the atoms in the ground state, while the ones in $(h_i)$ are expelled from the traps. We then turn on the fluorescence beams and image the recaptured atoms. This procedure maps the $\ket{\uparrow}$ and $\ket{\downarrow}$ states to the presence or absence of the corresponding atom.

When measuring spins in bases other than the natural $z$-basis, we apply a microwave pulse resonant with the $\ket{\uparrow} \leftrightarrow \ket{\downarrow}$ transition to rotate the spins prior to this readout sequence. The phase and pulse duration of the pulse are calibrated to match the target measurement basis.

The experimental sequence is typically repeated $\sim 1000~$times with at most $3~$ defects allowed in the $N = 114~$ assembled kagome arrays. 
This allows us to compute the magnetization and the spin correlations function by averaging over these repeated measurements.

\subsection{Addressing beams}\label{SubSM:addressing}

The addressing laser beams are generated by an external cavity, $1013~$nm diode laser seeding an amplifier outputting up to $8~$W. The light is blue detuned from the $(6P_{3/2}, m_J=3/2) \leftrightarrow \ket{\uparrow}$ transition by $\Delta/2\pi \sim 200~$MHz. We use a second dedicated SLM to produce the pattern of addressing beams, superimposed onto the tweezer array pattern. Each beam is focused on a $1/e^2$ radius of $1.5~\mu$m with a power of $\approx 50~$mW per atom resulting in a lightshift of $\delta_0/(2\pi)\approx 10$\, MHz.

\subsection{Experimental imperfections}\label{SubSM:exp_imperfection}

The experimental protocols outlined above are subject to various imperfections. 
We begin by examining errors related to state preparation and measurement (SPAM), and subsequently discuss the effects of decoherence in the system.

\subsubsection{Preparation and detection errors}\label{SubSubSM:prep_and_detec_errors}

\begin{figure}
	\includegraphics{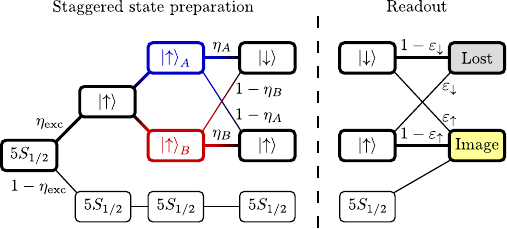}
\caption{\textbf{State preparation and mesurement error tree.} Left part: summarizes the different errors during the staggered state preparation. Right part: summarizes the different errors during the readout sequence. }
\label{fig:error_tree}
\end{figure}

Taking all these errors into account in a simulation is intractable since it would require simulating the XY dynamics of the $N=114$ atom system~\cite{Chen2023}. To overcome this issue, we create a simple model to describe the errors that can be implemented in a MontCarlo simulation. To do so, we divide the sequence into multiple steps, each having a probability of success and fail (see the error tree in Fig.~\ref{fig:error_tree}). During Rydberg excitation, $1-\eta_{\textrm{exc}}$ of the atoms remain in the ground state. Then, we apply the microwave pulses and the addressing. We denote $\eta_{\textrm{A}}$ and $\eta_{\textrm{B}}$ the probability to prepare, from $\ket{\uparrow}$, the addressed and non-addressed atoms in $\ket{\downarrow}$ and $\ket{\uparrow}$. Then we readout the state with detection errors of $\varepsilon_{\uparrow}$ and $\varepsilon_{\downarrow}$. Table~\ref{tab:error_tree_table} summarizes the different error sources, their physical origins, and their values, which are either measured or inferred from a series of dedicated experiments, or estimated from numerical simulations.

For atoms in sublattice $A$ (addressed), the error tree leads to the probability to recapture the atoms at the end of the staggered state preparation, which reads (to first order):
\begin{equation}
	p_A = (1-\eta_A) + (1-\eta_{\rm exc}) + \varepsilon_\uparrow
	\label{eq:P^z_A}
\end{equation}
Similarly, the calculation for sublattice $B$ (addressed atoms) yields the following:
\begin{equation}
	p_B = 1 - (1-\eta_B) - \varepsilon_\downarrow
	\label{eq:P^z_B}
\end{equation}
Using the values reported in Table~\ref{tab:error_tree_table}, we obtain $p_A = 6.6\%$, $p_B = 89.7\%$ that allows us to compute $P_z = p_B - p_A = 83.1\%$. We checked that these values agree with the measurement of the staggered magnetization at $t=0$ [see Fig.~\ref{fig:fig3}(c)], which is used as a calibration of the errors.

\renewcommand\arraystretch{1.0}
\setlength{\tabcolsep}{2pt}
\begin{table}
	\centering
	\begin{tabular}{|c|c|c|c|}
		\hline
		\textbf{Step} & \textbf{Symbol} & \textbf{Value} & \textbf{Main physical origin} \\
		\hline\hline

        \begin{tabular}{@{}c@{}} Rydberg \\ excitation \end{tabular}
        & $\eta_{\textrm{exc}}$ & $98(1)\%$
        &
        \begin{tabular}{@{}c@{}}
            Imperfect optical pumping,\\
            Laser phase noise,\\
            Spontaneous emission\\[-0.4ex]
            \hspace{1em}from $\ket{i}$~\citep{deleseleucAnalysisImperfectionsCoherent2018}
        \end{tabular} \\
		\hline	

		\multirow{2}{*}{\begin{tabular}{@{}c@{}} MW and \\ addressing \\ pulses \end{tabular}}
        & $\eta_B$ & $91(1)\,\%$
        &
        \begin{tabular}{@{}c@{}}
            Effect of the XY interactions\\
            during the MW pulses
        \end{tabular} \\
        \cline{2-4}
		& $\eta_A$ & $98(1)\,\%$
        &
        \begin{tabular}{@{}c@{}}
            Effect of the XY interactions\\
            during the MW pulse,\\
            Finite value of $\delta_0$
        \end{tabular} \\
		\hline

		\multirow{2}{*}{Read-out}
        & $\varepsilon_\downarrow$ & $1.3(2)\,\%$
        &
        \begin{tabular}{@{}c@{}}
            Hydrogenic state\\
            radiative lifetime\\
            Imaging infidelity\\
            (false negative)
        \end{tabular} \\
        \cline{2-4}
		& $\varepsilon_\uparrow$ & $2.6(2)\,\%$
        &
        \begin{tabular}{@{}c@{}}
            Imperfect MW freezing\\
            pulse,\\
            Imperfect deexcitation,\\
            Background gas collision\\
            losses,\\
            Imaging infidelity\\
            (false positive)
        \end{tabular} \\
		\hline		
	\end{tabular}

	\caption{\textbf{Summary of the experimental errors defined in Fig.~\ref{fig:error_tree}.} }
	\label{tab:error_tree_table}
\end{table}

\subsubsection{Sources of decoherence}\label{SubSubSM:decoherences}

In addition to the state preparation and measurement errors discussed above, two other mechanisms contribute to decoherence in our system.

{\it Positional disorder} — The finite spatial extent of the atomic wave packets in the traps, together with residual atomic motion, leads to shot-to-shot variations in the distances between atoms. In our model, both atomic positions and velocities are assumed to follow Gaussian distributions, characterized by standard deviations of 0.1~$\mu$m 
within the plane of the array, 
0.6~$\mu$m 
along the transverse direction, and 
0.03~$\mu$m$/\mu$s 
for all velocity components. These fluctuations in atomic separation result in variations of the interaction energy from one experimental realization to another, which give rise to decoherence when averaging over many shots.

{\it Finite Rydberg lifetime} — During the evolution, atoms excited to Rydberg states may either decay spontaneously back to the ground state at a rate $\Gamma^{\mathrm{sp}}$ or be transferred to other Rydberg levels through black-body radiation at a rate $\Gamma^{\mathrm{bb}}$. Such processes can bias the qubit-state readout and remove population from the computational basis, thereby modifying the subsequent dynamics. Using the values reported in Ref.~\cite{beterovQuasiclassicalCalculationsBlackbodyradiationinduced2009} for a temperature of $T = 300~\mathrm{K}$, we obtain the lifetimes $1/\Gamma^{\mathrm{sp}}_{60S} \simeq 260~\mu\mathrm{s}$, $1/\Gamma^{\mathrm{bb}}_{60S} \simeq 157~\mu\mathrm{s}$, $1/\Gamma^{\mathrm{sp}}_{60P} \simeq 472~\mu\mathrm{s}$, and $1/\Gamma^{\mathrm{bb}}_{60P} \simeq 161~\mu\mathrm{s}$.

\subsection{Kagome Brillouin zones}\label{app:BZ}
Here we summarize our conventions for the kagome lattice  geometry in real and momentum space.
First, we label the three interlacing triangular sublattices as $\mathfrak{a}$, $\mathfrak{b}$, and $\mathfrak{c}$ [see Fig.~\ref{fig:fig4_SM}].
We take a unit system where the spacing between nearest-neighbors is $a=1$, the positions within the unit cell are $\mb{\rho}_{\mathfrak{a}} = (0,0)$, $\mb{\rho}_{\mathfrak{b}} = (0,1)$,  and  $\mb{\rho}_{\mathfrak{c}}=(\sqrt{3}/2, 1/2)$,  and the basis vectors are $\mb{a}_1 = (\sqrt{3}, 1)$, $\mb{a}_2 = (0,2)$.
The set of all lattice sites is then  $\mathbf{r}_i = m_i \mb{a}_ + n_i \mb{a}_2 +\rho_i$, where $(m_i,n_i)\in\mathbb{Z}^2$.
Thus, the reciprocal vectors are  $\mb{b}_1 = (2\pi/\sqrt{3}, 0)$ and $\mb{b}_2 =(\pi, -\pi/\sqrt{3})$, which satisfy $\mb{a}_i \cdot \mb{b}_j  = 2\pi \delta_{ij}$. 
These generate a triangular Bravais lattice, around which momentum space can be divided into hexagonal Brillouin zones (BZ), depicted by the dash-dot hexagon in Fig.~\ref{fig:fig2}.
The high-symmetry points of this \textit{inner}  BZ are $\Gamma=(0,0)$,  $M_I^{(i)} = \pm \pi \cdot (\cos(2\pi i/6),\,\sin(2\pi i/6))$, where $i=1,2,3$; and $K_I^\pm = \pm (\pi, \pi/\sqrt{3})$. 
Additionally, we can consider the \textit{extended} BZ, depicted as the dashed hexagon in Fig.~\ref{fig:fig2}, which naturally arises when computing the spin structure factor.
It is the BZ of the finer triangular lattice formed by adding a site to the center of each hexagon.
The extended BZ is twice the linear size of the inner BZ, and we label its high symmetry points as $M^{(i)} = 2 M_I^{(i)}$ and $K^\pm = 2 K_I^\pm$.

\subsection{Further analysis of the correlations}\label{SubSM:correlation}

\begin{figure*}
	\centering
	\includegraphics
    {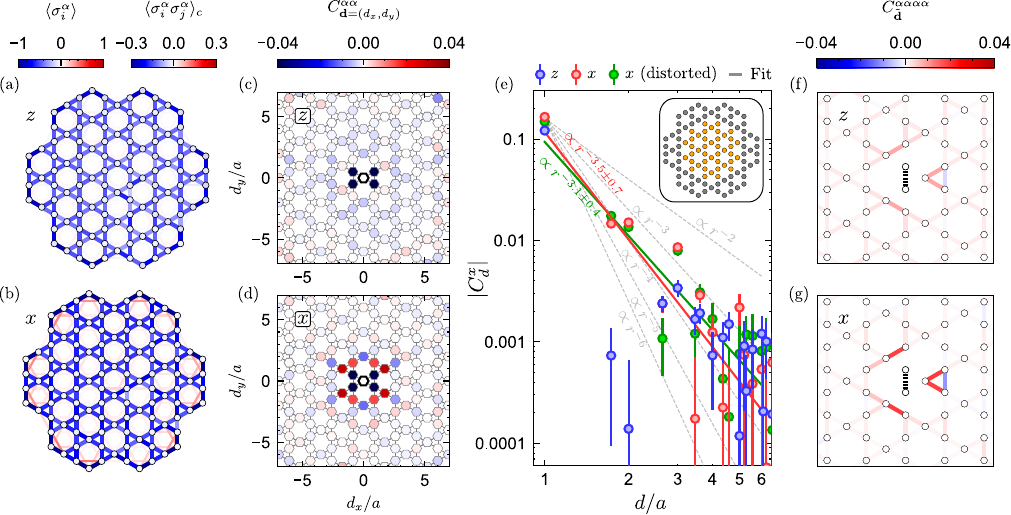}	\caption{\textbf{Detailed analysis of the magnetizations and correlations measured at the end of the adiabatic ramp.}  The first row displays measurements along the spin axis $\alpha=z$, and the second along $\alpha=x$.
        (a)/(b) represent the magnetization $\langle \sigma^{\alpha}_i \rangle$ of each atom (colored circles) and the connected nearest-neighbor and next-nearest-neighbor correlations $\langle \sigma^{\alpha}_i\sigma^{\alpha}_j \rangle_c$ (colored bonds). 
        (c)/(d) Map of correlation $C^{\alpha\alpha}_{\mathbf{d} = (d_x, d_y)}$.
        (g) Averages correlation $|C^{\alpha\alpha}_{d}|$ as a function of the distance $d$. The inset represents the 42 atoms in the bulk of array (orange site) used to calculate all these correlation functions. 
        (e)/(f) Four body connected correlation $C^{\alpha\alpha\alpha\alpha}_{\tilde{\mathbf{d}}}$.}
	\label{fig:fig3_SM}
\end{figure*}

Figure~\ref{fig:fig3_SM} shows a more detailed analysis of the correlations measured at the end of the adiabatic procedure. Figures~\ref{fig:fig3_SM}(a) and (b) display the single-atom magnetization along the $z$ and $x$ directions (colored circles), together with the nearest- and next-nearest-neighbor correlations along the same axes, represented by colored bonds. At the end of the adiabatic ramp, the system has lost its staggered magnetization, and the atoms exhibit vanishing local magnetization in both measurement bases. In contrast, we observe relatively strong antiferromagnetic nearest-neighbor correlations along both the $z$ and $x$ directions.

\begin{figure}
	\centering
	\includegraphics
    {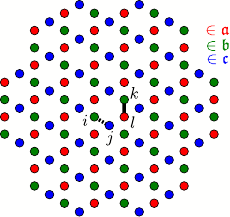}	\caption{
    \textbf{Sublattice structure and correlation analysis.} Each site of the kagome lattice belongs to one of three sublattices ($a$, $b$, $c$), which define the orientation of nearest-neighbor bonds. For the two-body correlation map, each pair $(i,j)$ is rotated according to the sublattice of atom $j$ to define a pair-frame displacement $R(\theta_j)\mathbf{r}_{ij}$, which is then used to average correlations over geometrically equivalent pairs. For the four-body connected correlations, a reference nearest-neighbor pair $(i,j)$ (black dashed bond) is translated and rotated to a canonical configuration, and a second nearest-neighbor pair $(k,l)$ (red bond) is characterized by the displacement $\tilde{\mathbf{d}}_{kl}$ between the midpoints of the two bonds.
    }
	\label{fig:fig4_SM}
\end{figure}

To analyze the spatial structure of the two-point correlations in a manner independent of the global lattice orientation, we construct maps by exploiting the threefold rotational symmetry of the kagome lattice. For each atom pair $(i,j)$, we define the relative displacement vector $\mathbf{r}_{ij} = \mathbf{r}_i - \mathbf{r}_j$. Each lattice site belongs to one of the three kagome sublattices, labeled $\mathfrak{a}$, $\mathfrak{b}$, and $\mathfrak{c}$ [see Fig.~\ref{fig:fig4_SM}]. Depending on the sublattice of atom $j$, we associate a rotation angle $\theta_j$ taking the values $+\pi/3$, $-\pi/3$, or $0$. We then define a \emph{pair-frame displacement} $R(\theta_j)\,\mathbf{r}_{ij}$, which maps geometrically equivalent pairs onto a common reference orientation.

Using these pair-frame displacements, we define the correlation map
\begin{equation}
	C^{\alpha\alpha}_{\mathbf{d}} = \frac{1}{N_\mathbf{d}} 
	\sum_{\{(i,j)\mid R(\theta_j)\,\mathbf{r}_{ij}=\mathbf{d}\}} 
	\langle \sigma^{\alpha}_i \sigma^{\alpha}_j \rangle_c ,
	\label{eq:corr_map}
\end{equation}
where the sum runs over all atom pairs whose pair-frame displacement equals $\mathbf{d}$, and $N_\mathbf{d}$ denotes the corresponding number of pairs. To reduce edge effects, only pairs for which atom $i$ is located within the central 42 sites of the array [highlighted in orange in the inset of Fig.~\ref{fig:fig3_SM}(e)] are included, while atom $j$ may lie anywhere in the array.

Figures~\ref{fig:fig3_SM}(c) and (d) show the resulting correlation maps measured along the $z$ and $x$ directions at the end of the ramp. From these maps, we further compute the distance-dependent average of the spin correlator,
\begin{equation}
	|C^{\alpha\alpha}_{d}| = \frac{1}{N_d} 
	\sum_{\{(i,j)\mid |\mathbf{r}_{ij}|=d\}} 
	\langle \sigma^{\alpha}_i \sigma^{\alpha}_j \rangle_c ,
	\label{eq:corr_map_dist}
\end{equation}
where the sum is taken over the same subset of atom pairs as in Eq.~\eqref{eq:corr_map}, now grouped according to the magnitude of their displacement, and $N_d$ denotes the number of such pairs. The results are shown in Fig.~\ref{fig:fig3_SM}(e) as a function of the interatomic distance $d$, including correlations measured along the $z$ direction (blue), along the $x$ direction (red), and along the $x$ direction for the distorted array (pink; see main text). The $z$ correlations exhibit a nontrivial, non-monotonic decay with distance, whereas the correlations measured along $x$ display an approximately algebraic decay with similar power-law exponents (from fits to $|C^x_d|\propto r^{-\alpha}$) for both the distorted and undistorted arrays.

Finally, we compute the connected four-body correlations for quadruplets of atoms formed by two nearest-neighbor pairs $(i,j)$ and $(k,l)$. As for the two-point correlations, we analyze these quantities in a reference frame independent of the global lattice orientation. We first translate the coordinates such that the midpoint of the reference pair $(i,j)$ lies at the origin. Nearest-neighbor bonds on the kagome lattice fall into three classes, corresponding to the sublattice pairs $\mathfrak{a}\!-\!\mathfrak{b}$, $\mathfrak{b}\!-\!\mathfrak{c}$, and $\mathfrak{c}\!-\!\mathfrak{a}$, each of which uniquely defines a bond orientation in real space [see Fig.~\ref{fig:fig4_SM}]. We apply a rotation determined by the sublattice class of the pair $(i,j)$ so that all reference bonds are aligned along the same direction, thereby mapping all geometrically equivalent configurations onto a common canonical frame.

After this transformation, the positions of atoms $k$ and $l$ are denoted by $\tilde{\mathbf{r}}_{k}$ and $\tilde{\mathbf{r}}_{l}$. We characterize the relative position of the two nearest-neighbor pairs by the displacement between their midpoints, $\tilde{\mathbf{d}}_{kl} = (\tilde{\mathbf{r}}_{k} + \tilde{\mathbf{r}}_{l})/2$. Because the reference bond $(i,j)$ is fixed and $(k,l)$ is restricted to nearest neighbors, the lattice geometry uniquely determines the orientation of the $(k,l)$ bond for a given displacement $\tilde{\mathbf{d}}_{kl}$. As a result, the geometry of each quadruplet is fully specified by $\tilde{\mathbf{d}}_{kl}$.

We then average the four-body connected correlation over all quadruplets with the same displacement $\tilde{\mathbf{d}}$ as
\begin{equation}
C^{\alpha\alpha\alpha\alpha}_{\tilde{\mathbf{d}}} = \frac{1}{N_{\tilde{\mathbf{d}}}} \sum_{\substack{(i,j,k,l) \\
\tilde{\mathbf{d}}_{kl}=\tilde{\mathbf{d}} \\
|\mathbf{r}_{ij}| = |\mathbf{r}_{kl}| = a \\
i \neq j \neq k \neq l
}} \langle \sigma^\alpha_i \sigma^\alpha_j \sigma^\alpha_k \sigma^\alpha_l \rangle_c ,
\label{eq:4_body_corr_average}
\end{equation}
where $N_{\tilde{\mathbf{d}}}$ is the number of quadruplets contributing to a given displacement. To avoid edge effect, the sum takes $(i,j)$ to be in the center 42 sites [orange sites shown in the inset of Fig.~\ref{fig:fig3_SM}(e)], while $(k,l)$ is allowed to range over the full array. The four-body connected correlation are defined following the Ursell functions~\cite{kuboGeneralizedCumulantExpansion1962}:
\begin{equation}
\begin{aligned}
\langle AB \rangle_c &= \langle AB \rangle - \langle A \rangle \langle B \rangle, \\[2mm]
\langle ABC \rangle_c &= \langle ABC \rangle
- \langle AB \rangle_c \langle C \rangle
- \langle AC \rangle_c \langle B \rangle \\
&\quad - \langle BC \rangle_c \langle A \rangle
- \langle A \rangle \langle B \rangle \langle C \rangle, \\[2mm]
\langle ABCD \rangle_c &=
\langle ABCD \rangle
- \langle ABC \rangle_c \langle D \rangle
- \langle ABD \rangle_c \langle C \rangle \\
&\quad- \langle ACD \rangle_c \langle B \rangle
- \langle BCD \rangle_c \langle A \rangle \nonumber \\
&\quad- \langle AB \rangle_c \langle CD \rangle_c
- \langle AC \rangle_c \langle BD \rangle_c \\
&\quad- \langle AD \rangle_c \langle BC \rangle_c \nonumber 
- \langle AB \rangle_c \langle C \rangle \langle D \rangle \\
&\quad- \langle AC \rangle_c \langle B \rangle \langle D \rangle
- \langle AD \rangle_c \langle B \rangle \langle C \rangle \nonumber \\
&\quad- \langle BC \rangle_c \langle A \rangle \langle D \rangle
- \langle BD \rangle_c \langle A \rangle \langle C \rangle\\
&\quad- \langle CD \rangle_c \langle A \rangle \langle B \rangle \nonumber
- \langle A \rangle \langle B \rangle \langle C \rangle \langle D \rangle .
\end{aligned}
\label{eq:4_body_corr_Ursell}
\end{equation}
with $A=\sigma^\alpha_i$, $B=\sigma^\alpha_j$, $C=\sigma^\alpha_k$, and $D=\sigma^\alpha_l$.

\subsection{DSL Ansatz construction}\label{SubSM:Ansatz}

\begin{figure}
	\centering
	\includegraphics
    {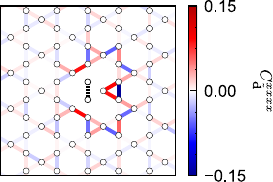}	\caption{
    \textbf{Four body connected correlation $C^{xxxx}_{\tilde{\mathbf{d}}}$ of the DSL ansatz wavefunction.}  The sign structure, i.e. of the nearest five bonds, matches that seen in the experiment [Fig.~\ref{fig:fig3_SM}(g)], although the magnitude is larger.
    }
	\label{fig:fig6_SM}
\end{figure}

The $U(1)$ DSL \textit{ansatz} is constructed in a standard way, as follows~\cite{ranProjectedWaveFunctionStudySpin2007, beccaQuantumMonteCarlo2017}.
First, we suppose the low-energy excitations are emergent fermions with a Dirac cone band structure.
Microscopically, their mean-field kinetic energy is given by a nearest-neighbor hopping Hamiltonian, $H_{\rm{MF}} = \sum_{i,j,\sigma}  c^\dagger_{i,\sigma} e^{i\theta_{ij}}  c_{j,\sigma}$,
where $c^\dagger, c$ are fermionic ladder operators, $\sigma$ is the spin index, and $\theta_{ij}\in \mathbb{R}/2\pi$.
In the thermodynamic limit, $\theta_{ij} \in \{0,\pi\}$, so that moving around a kagome triangle encloses $0$ flux, and around a hexagon encloses $\pi$ flux.
When both spin species of fermions are brought to half-filling, the lowest energy state is a Slater determinant $|\psi_{\rm{MF}}\rangle$ of single particle eigenstates, with on average one fermion per site, $\langle \psi_{\rm{MF}}|c_{i,\uparrow}^\dagger c_{i,\uparrow} + c_{i,\downarrow}^\dagger c_{i,\downarrow}|\psi_{\rm{MF}}\rangle =1$.
Enforcing this single-occupation to hold exactly by a Gutzwiller projection, $\Pi_G = \prod_i (1-n_{i,\uparrow}n_{i,\downarrow})$, yields the \textit{ansatz} spin wavefunction, $\ket{\psi_G} = \Pi_G\ket{\psi_{\rm{MF}}}$.
The properties of this state can be efficiently determined by Monte Carlo sampling.
This includes, for instance, multi-body spin correlation functions, such as the four-body correlation $C^{xxxx}_{\tilde{\mathbf{d}}}$ we measured in the experiment; the \textit{ansatz} prediction for this observable is shown in Fig.~\ref{fig:fig6_SM}.

One subtlety for finite-size systems is that boundary conditions can lead to a mean-field fermionic degeneracy at half-filling.
This is the case for our $N=114$ cluster with exactly $\pi$ flux per hexagon, and so to have a uniquely-defined Slater determinant we must weakly modify the gauge angles $\theta_{ij}$ in $H_{\rm{MF}}$.
We choose to uniformly spread a single quantized flux $\Phi_{\rm{tot}}=2\pi$ across the system, so that each lattice hexagon encloses $\pi+2\pi/31$ magnetic gauge flux. 
Doing so gaps the degeneracy and gives a well-defined \textit{ansatz} wavefunction, at the expense of small chiral symmetry breaking.
This is a finite-size effect that vanishes for larger systems.

To calculate the Friedel response shown in in Fig.~\ref{fig:static}(e), we perturb the mean-field Hamiltonian with a local magnetic field for the fermionic spinons,
\begin{equation}
    H_{\rm{MF}} = \left(t \sum_{i,j,\sigma}  c^\dagger_{i,\sigma} e^{i\theta_{ij}}  c_{j,\sigma}\right) + h (c_{0,\uparrow}^\dagger c_{0,\uparrow} - c^\dagger_{0,\downarrow} c_{0,\downarrow})
\end{equation}
where $t$ is the (uniform) spinon hopping amplitude, and $h$ is the field strength.
We then construct the corresponding Slater determinant and its Gutzwiller projection, from which we measure the local  expectation value $\langle \sigma^z_i\rangle$ as function of $h/t$.
The $(h/t)$-dependence of $\sigma^z_i$ is linear at small-field, so these slopes define a response function $\chi_i$.
We caution that there is an overall factor undetermined in this approach, namely the relation between the mean-field $h/t$ and the physical $\delta/J$.
Still, the relative response between different sites, $\chi_i/\chi_j$, is meaningful.

\subsection{Correlation functions for competing states}

Kagome lattice antiferromagnet phase diagrams often feature many ground state phases that compete at low energy scales.
Here, we consider two important ones that are closely related to the $U(1)$ DSL.
For each, we take a simple \textit{ansatz} state representative of that phase on the $N=114$ cluster, and compute its spin-spin correlations analogous to those shown for the experiment and the $U(1)$ DSL in main text Fig.~\ref{fig:ansatz}.
%

\begin{figure}
	\centering
	\includegraphics
    {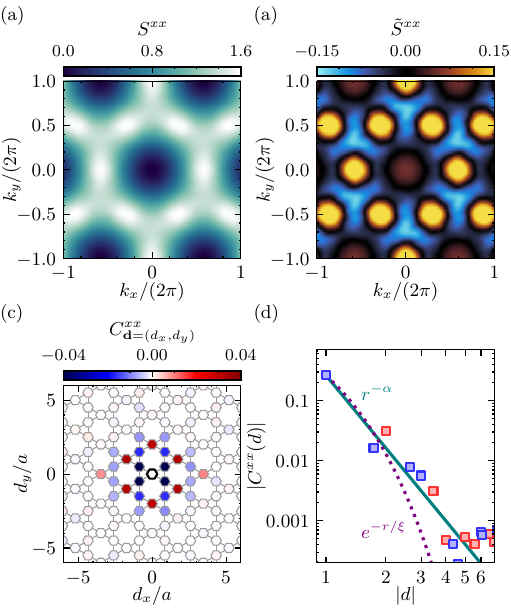}	\caption{\textbf{Correlations of the $Z_2[0,\pi]\beta$ spin liquid.}
    (a) Spin structure factor. (b) Modified structure factor (nearest-neighbor correlations subtracted). (c) Real-space correlations as a function of displacement. (d) Spatial decay of correlation magnitudes (red markers for $C^{xx}>0$, blue markers for $C^{xx}<0$).  In each plot, we average over pairs $(i,j)$ with $i$ in the center 42 sites, and $j$ anywhere in the array.  Best-fits are $\alpha=4.02(1)$ for a power-law, and $\xi=0.33(5)$ for an exponential decay.
		}
	\label{fig:SM_Z2}
\end{figure}

%
First, we consider a gapped spin liquid with $Z_2$ topological order, and symmetry fractionalization pattern labelled as $Z_2[0,\pi]\beta$ in Ref.~\cite{lu2SpinLiquids2011}.
This phase is obtained from the U(1) DSL via a Higgs transition driven by the pairing of the fermionic spinons, and is known to have very low energies for the nearest-neighbor Heisenberg antiferromagnet.
As a representative, we choose a particular point in this phase where the ground state is a signed superposition of nearest-neighbor valence bond coverings of the kagome lattice~\cite{yangFrustratedResonatingValence2012, luUnificationBosonicFermionic2017}.
The spin-spin correlations are computed by a similar Monte Carlo sampling as we use for the U(1) DSL, and are shown in Fig.~\ref{fig:SM_Z2}.
Overall, the momentum-space [Fig.~\ref{fig:SM_Z2}(a,b)] and real-space [Fig.~\ref{fig:SM_Z2}(c,d)] correlations at this system size are very similar to those for the $U(1)$ DSL (main text Fig.~\ref{fig:fig3}).
One short-range distinction seems to be the next-nearest-neighbor ($d/a=\sqrt{3}$) correlations, which are negative in this $Z_2[0,\pi]\beta$ state but weakly positive in the $U(1)$ DSL ansatz.
Second, we examine a conventional, $U(1)$-symmetry-breaking ordered phase: the coplanar antiferromagnet with 120$^\circ$ oriented spins and a three-site unit cell (i.e. ordering wavevector $q=0$).
This ordered state descends from the $U(1)$ DSL through proliferation of spin-triplet magnetic monopoles~\cite{songUnifyingDescriptionCompeting2019}.
%

\begin{figure}
	\centering
	\includegraphics
    {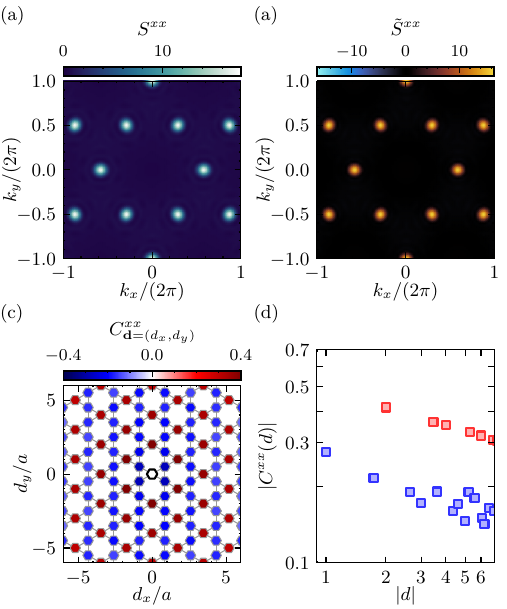}	\caption{\textbf{Correlations of the $q=0$ coplanar magnetic order.}     (a) Spin structure factor. (b) Modified structure factor (nearest-neighbor correlations subtracted). (c) Real-space correlations as a function of displacement. (d) Spatial decay of correlation magnitudes (red markers for $C^{xx}>0$, blue markers for $C^{xx}<0$).  In each plot, we average over pairs $(i,j)$ with $i$ in the center 42 sites, and $j$ anywhere in the array.
		}
	\label{fig:SM_LSWT}
\end{figure}

%
We compute the correlations for this phase under the approximations of linear spin wave theory~\cite{peterAnomalousBehaviorSpin2012}, with some methodological adaptations beyond the textbook case needed for our finite size system.
In particular, we begin by numerically finding a local minimum of the mean-field energy (which accounts for edge effects) near the $q=0$ ordered configuration, giving a local $xy$ spin orientation $\{\eta_i^0\}\in[0,2\pi)^N$.
We then perform a global rotation so that each spin lies along the $x$ direction, and then in this rotated frame compute the spin length $m_i=\langle \tilde{S}^x_i\rangle$ and the correlation function of transverse fluctuations, $\langle \delta \tilde{S}^y_i \delta \tilde{S}^y_j\rangle $.
This step is done using the Cholesky decomposition method described by Colpa~\cite{colpaDiagonalizationQuadraticBoson1978}.
Finally, we obtain the physical correlation function by rotating back to the lab frame, and then averaging over the global $U(1)$ symmetry breaking direction, which yields 
\begin{equation}
        C^{xx}_{ij} = 4\cos(\eta_i-\eta_j)
        \times\left[ m_im_j + \langle \delta \tilde{S}^y_i \delta \tilde{S}^y_j\rangle\right].
\end{equation}
The results (Fig.~\ref{fig:SM_LSWT}) show that, even for this system size, spin wave theory predicts strong off-diagonal long-range order, i.e. at long-distances $C^{xx}(d)\approx 0.3$ for positive, same-sublattice correlations, and $C^{xx} \approx -0.15$ for negative, different-sublattice correlations.

\subsection{Additional comparisons between experiment and ansatz}\label{SubSM:Exp_vs_Ga}

\begin{figure}
	\centering
	\includegraphics
    {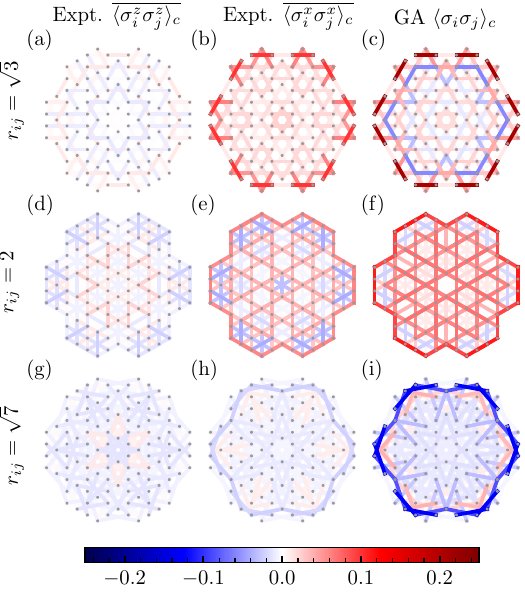}	\caption{\textbf{Radial dependence of correlations.}  Each panel shows local correlations $\langle \sigma_i\sigma_j\rangle_c$, for pairs at separation $r_{ij}=(\sqrt{3},2,\sqrt{7})$ (top, middle, bottom rows).  Left, center, right columns correspond to experimental $z$-basis, experimental $x$-basis, and Gutzwiller ansatz (basis independent) correlations.  For the experimental data, we average over the $D_6$ spatial symmetry of the kagome lattice cluster, to highlight the radial dependence (i.e. bulk vs. boundary). 
		}
	\label{fig:SM_radial}
\end{figure}

One important aspect of our finite-size system is the presence of an edge, near which physical properties can generally differ from in the bulk.
As an example, in Fig.~\ref{fig:SM_radial} we display how the beyond-nearest-neighbor spin correlations vary across the system, comparing between the experiment and the DSL Gutzwiller ansatz.
We are primarily interested in the bulk-edge comparison, so we average over the $D_6$ spatial symmetry group of the kagome lattice cluster (i.e. reflections and rotations).
This removes angular inhomogeneities induced by experimental imperfections, leaving only the radial dependence---some of which should reflect intrinsic boundary physics.
Each panel of Fig.~\ref{fig:SM_radial} shows the two-body connected correlation function $\langle \sigma^\alpha_i \sigma^\alpha_j\rangle$ between pairs of atoms at separation $r_{ij}=(\sqrt{3},2,\sqrt{7})$, (top, middle, and bottom rows).
The first two columns show the experimental data from the end of the adiabatic ramp, along $\sigma^z$ and $\sigma^x$,  and the third column shows the prediction from the Gutzwiller ansatz (which is basis-independent).
Generally, we find that the experimental $\sigma^x$ and Gutzwiller ansatz correlations have some notable similarites, e.g. the enhanced positive correlations at $r_{ij}=\sqrt{3}$ near the edge. 
The experimental $\sigma^z$ correlations are weaker, and do not exhibit the same boundary enhancement. 

\end{document}